# *Anisotropic photonic time interfaces via isotropic spacetime modulations*

*Andrew M. Naylor[1] and Victor Pacheco-Peña[1*]*

[1] *School of Mathematics, Statistics and Physics, Newcastle University, Newcastle Upon Tyne, NE1 7RU, United Kingdom*
*email: victor.pacheco-pena@newcastle.ac.uk*

**The engineering of the optical properties of materials in space and time is opening further directions and possibilities to control wave propagation in four dimensions (*x,y,z,t*). A key example of such modulations are time interfaces where the permittivity of the medium is changed in time from isotropic to another isotropic value. Recently, isotropic-to-anisotropic time interfaces in a homogeneous, unbounded medium have also been proposed, demonstrating their potential for redirecting waves in real time. However, the challenge relies on accessing/creating permittivity tensors in time. To address this, here we propose isotropic-to-isotropic spacetime modulations inspired by spacetime effective media to emulate such isotropic-to-anisotropic time interfaces. Specifically, we consider that subwavelength spatially periodic multilayers, arranged either horizontally or vertically, are created in time using simultaneous isotropic-to-isotropic time interfaces applied in discrete spatial regions. The theory behind this approach is presented in detail demonstrating that, indeed, it is possible to change the direction of energy propagation in real time and emulate permittivity tensors. All the results are supported by numerical simulations, demonstrating the potential of the proposed spacetime approach to emulate photonics isotropic-to-anisotropic time interfaces.**

# 1. Introduction

Manipulating wave propagation in space $(x, y, z)$ has been at the centre of multiple fields including microwave engineering, photonics, plasmonics and metamaterials [1–5]. Research and innovation in these areas have enabled the improvement of known technologies as well as the proposal of new applications with examples including sensing[6–11], computing[12–29] and antennas[30–38], to name a few[39–44]. In seeking further ways to tailor wave-matter interactions, there has been a growing interest into increasing the available degrees of freedom by bringing together space and time ($t$). In this scenario, the optical properties of the materials (permittivity, $\varepsilon_r$, and/or permeability, $\mu_r$) where a wave propagates are no longer considered to be time-independent, but they can now vary both in space and time. This has opened further opportunities to tailor waves in four-dimensions (4D) $(x, y, z, t)$[45–54].

Among different types of temporal modulations of the optical properties of materials, time interfaces have particularly attracted the attention of the scientific community. Introduced by Morganthaler[55], they consist of a rapid variation in time (with a time duration smaller than the period $T$ of the incident signal) of the electromagnetic (EM) properties of the material where a wave propagates between an initial ($\varepsilon_{r1}$ and/or $\mu_{r1}$) and a final relative value ($\varepsilon_{r2}$ and/or $\mu_{r2}$). In so doing, a forward and a backward wave (FW and BW, respectively) can be produced, which are widely known as the temporal analogue of transmitted and reflected waves, respectively, generated when a wave encounters a sharp spatial interface between two materials with different impedances[56]. Research involving such time-modulated media has given rise to the study and proposal of interesting phenomena including inverse prism[57], effective medium theory in time and spacetime[49,58–61], beam splitting[62], spatiotemporal gratings[63], temporal anisotropy[64–70], quantum applications[71–73], frequency conversion[74,75], slow temporal modulations[76,77] and filters[78,79], among other examples[80–89]. Interestingly, the scientific community has also started to report experimental demonstrations, showing the interesting physical phenomena that can arise when manipulating wave in 4D[90–96].

In this realm, recent studies have shown that time interfaces involving isotropic-to-anisotropic modulations of the optical properties of materials can lead to interesting physics including the capability of enabling temporal beam steering (or temporal aiming)[70]. This is done by considering a $p$-polarized wave propagating within an unbounded isotropic medium having a relative permittivity $\varepsilon_{r1}$ for times $t < t_0$. If an isotropic-to-anisotropic time interface is induced at $t = t_0$, by changing the permittivity to a tensor (namely $\overline{\overline{\varepsilon_{r2}}} = \{\varepsilon_{r2x}, \varepsilon_{r2z}\}$, all values larger than unity), the direction of energy propagation (Poynting vector) can be modified in time, while the wavenumber, *k*, does not change (as required by a time interface[55]). Such interesting physical phenomena has also been extended to three-dimensional space[97] and demonstrated using water waves[69], showing the potential of translating such physical phenomena beyond photonics. While recent works have also proposed further interesting phenomena behind anisotropy in time modulated media[64–70], the question of how to access such isotropic-to-anisotropic time modulations in photonics requires further exploration.

Inspired by the importance of time interfaces and spacetime photonics, here we ask: would it be possible to create/emulate isotropic-to-anisotropic photonic time interfaces using purely isotropic-to-isotropic time interfaces? If this is possible then it could open potential opportunities for their experimental demonstration in photonics. Here, we answer this question by exploiting effective medium theory in space and time[58,59]. Specifically, we study the case where a wave propagates within an unbounded medium having an isotropic, homogeneous, relative permittivity, $\varepsilon_{r1}$, for times $t < t_0$. Then, at $t = t_0$ a temporal interface is introduced in certain regions in space in such a way that spatially subwavelength periodic multilayers are rapidly induced in time. An in-depth theoretical study is presented demonstrating how for times $t > t_0$ an effective anisotropic medium is created. Different configurations are studied including induced subwavelength spatial multilayers organized vertically (parallel to the *x* axis, considering propagation on the *xz* plane) or horizontally (parallel to the *z* axis). The direction of the energy propagation, Poynting vector, and amplitude of the generated FW and BW waves are theoretically calculated after inducing such effective isotropic-to-anisotropic time interface via isotropic-to-isotropic spacetime modulations. Further studies include different values of $\varepsilon_r$ for the temporally induced spatially subwavelength periodic layers as well as the effect of the filling fraction of each induced spatial layer. All the results are validated numerically using COMSOL Multiphysics®. These results may pave the way for potential experimental validations and further explorations of isotropic-to-anisotropic photonic time interfaces involving only spatiotemporal isotropic modulations.

## 2. Results and Discussion

### 2.1 Theory.

To begin with, let us first consider the scenario shown in Fig.1a-c. This case has been studied in detail in[70] but we will briefly summarize it here for completeness. As observed in Fig. 1a, an "oblique" *p*-polarised EM wave is propagating within an isotropic, unbounded medium (relative permeability and permittivity, $\mu_{r1}$ and $\varepsilon_{r1}$, respectively) for times $t < t_0$. The incident angle of the EM wave is $\theta_1$ measured on the $xz$ plane and the wavenumber, $\boldsymbol{k}$, and Poynting vector, $\boldsymbol{S}$, are parallel such that $\theta_1 = \theta_{1k} = \theta_{1S}$, as expected. Note that here, even when the angle of incidence is "oblique", for times $t < t_0$ the medium is isotropic and unbounded meaning that such angle is, strictly speaking, irrelevant. However, it becomes important when an anisotropic time interface is introduced in time, as follows: as shown on the top panels from Fig. 1b,c, at $t = t_0$ an isotropic-to-anisotropic time interface is induced everywhere in space such that the relative permittivity is changed to a tensor (i.e., $\overline{\overline{\varepsilon_{r2}}} = \{\varepsilon_{r2x}, \varepsilon_{r2z}\}$, $\varepsilon_{r2x}$ and $\varepsilon_{r2z}$ as the relative values along the $x$ and $z$ axis, respectively; all values greater than unity). Moreover, as shown in Fig. 1a-c, as the magnetic field is out-of-plane, changing the permeability will only account for an isotropic time interface. Therefore, as in [66,70], $\mu_r$ can be assumed to be time-independent (or $\mu_{r1} = \mu_{r2y}$) (the full expressions considering time-dependent $\mu_r$ are shown in the Supplementary materials document). With this configuration, the schematic representations of the produced FW and BW waves due to the time interface along with the direction of $\boldsymbol{S}$ and $\boldsymbol{k}$ for both waves are presented in Fig. 1b,c for the cases where $\varepsilon_{r2x} < \varepsilon_{r2z}$ and $\varepsilon_{r2x} > \varepsilon_{r2z}$, respectively. As reported before, and explained in the introduction, the FW and BW waves created by such isotropic-to-anisotropic time interfaces preserve $\boldsymbol{k}$ whereas the direction of energy propagation, $\boldsymbol{S}$, is modified ($\theta_1 = \theta_{2k} \neq \theta_{2S}$), meaning that waves can be steered in real time to angles that are larger or smaller than the initial angle for times $t < t_0$[69].

Now, the scenario described in Fig. 1a-c describes the case where the isotropic-to-anisotropic time interface is applied everywhere in space (homogeneous, unbounded medium). The question here is, how could we recreate such a photonic time interface? Would it be possible not to rely on a permittivity tensor but to use isotropic-to-isotropic spacetime modulations instead to recreate the performance shown in Fig. 1a-c? To answer this question, here we propose the scenario shown in Fig. 1d-f. First, we consider that for times $t < t_0$ we have the same scenario as described in Fig. 1a (i.e., an "oblique" incident wave propagating within an isotropic, unbounded medium). Then, at $t = t_0$, multiple isotropic-to-isotropic time interfaces are induced at discrete spatial regions, resulting in the formation of a spatially subwavelength multilayered structure (i.e., an isotropic-to-isotropic spacetime modulation). As shown in Fig. 1d-f, we will consider two cases: when "horizontal" (parallel to the *z*-axis) and "vertical" (parallel to the *x*-axis) subwavelength periodic multilayers are induced simultaneously in time. Interestingly, the spacetime effective medium theory arising from such configuration has been recently explored[59] considering normal incidence (i.e., either $\theta_1 = 0°$ or $\theta_1 = 90°$ for the horizontal or vertical subwavelength multilayers, respectively). In that work, it was shown that one could describe the structure via an effective relative permittivity tensor with $\varepsilon_{reff,x} \neq \varepsilon_{reff,z}$ (as the relative effective values along the *x* and *z* axes, respectively) which will vary depending if the multilayers are induced vertically or horizontally. In our present work, however, we are interested in studying the scenarios from Fig. 1d-f where the angle of the signal for $t < t_0$ is "oblique". This means that, when the horizontal or vertical subwavelength periodic multilayers are simultaneously induced via multiple isotropic-to-isotropic time interfaces at $t = t_0$ (Fig. 1e,f), the incident signal will no longer be just perpendicular or parallel to the spatial multilayers, respectively. As it will be shown later, we demonstrate that such scenario can indeed be used to mimic the performance of an isotropic-to-anisotropic interface without the need of anisotropic temporal modulations.

To begin with, let us discuss the scenario from Fig. 1e where horizontal subwavelength periodic multilayers are induced simultaneously in time at $t = t_0$. As observed in the figure, the periodic multilayered structure consists of two types of layers (namely layers A and B). For simplicity, and without loss of generality, we consider the case where the permittivity of layers A is time-independent ($\varepsilon_{rA} = \varepsilon_{r1}$) while for layers B $\varepsilon_{rB} \neq \varepsilon_{r1}$. As it is known[59], the effective permittivity tensor for times $t > t_0$ for this scenario is:

$$\varepsilon_{reff,z} = \widetilde{\Delta x_A}\varepsilon_{rA} + (1 - \widetilde{\Delta x_A})\varepsilon_{rB} \tag{1a}$$

$$\varepsilon_{reff,x} = \frac{\varepsilon_{rA}\varepsilon_{rB}}{(1 - \widetilde{\Delta x_A})\varepsilon_{rA} + \widetilde{\Delta x_A}\varepsilon_{rB}} \tag{1b}$$

where $\widetilde{\Delta x_A} = \Delta x_A/(\Delta x_A + \Delta x_B)$ is the filling fraction of layers A, and $\Delta x_{A,B}$ are the absolute thicknesses of layers A and B along the *x* axis, respectively, having isotropic relative permittivity values of $\varepsilon_{rA,B}$, respectively. Importantly, Eq. 1, which describes the scenario from Fig. 1e, takes the form of $\varepsilon_{reff,x} < \varepsilon_{reff,z}$ for all values of the isotropic relative permittivities of the induced spatial layers $\varepsilon_{rA}$ and $\varepsilon_{rB}$. Thus, this scenario can be understood as the isotropic spacetime equivalent of the case presented in Fig. 1b, where an anisotropic tensor was induced in time and everywhere in space. A similar approach can be followed for the vertical subwavelength periodic multilayers schematically shown in Fig.1f. Similar to Eq. 1, for this scenario the permittivity tensor can be described by using effective medium theory with the components defined as follows:

$$\varepsilon_{reff,z} = \frac{\varepsilon_{rA}\varepsilon_{rB}}{(1-\widetilde{\Delta z_A})\varepsilon_{rA} + \widetilde{\Delta z_A}\varepsilon_{rB}} \quad (2a)$$

$$\varepsilon_{reff,x} = \widetilde{\Delta z_A}\varepsilon_{rA} + (1-\widetilde{\Delta z_A})\varepsilon_{rB} \quad (2b)$$

where $\widetilde{\Delta z_A} = \Delta z_A/(\Delta z_A + \Delta z_B)$ is the filling fraction of layers A, and $\Delta z_{A,B}$ are the absolute thicknesses of layers A and B along the $z$ axis, respectively. As observed from Eq. 2, as the components of the effective relative permittivity tensor are exchanged (compared to the ones shown in Eq. 1 for the case when horizontal subwavelength periodic multilayers are induced in time), one will expect that now $\varepsilon_{reff,x} > \varepsilon_{reff,z}$. This means that the spacetime structure from Fig.1f is equivalent to the scenario shown in Fig.1c where an anisotropic tensor was induced in time and everywhere in space with $\varepsilon_{r2x} > \varepsilon_{r2z}$.

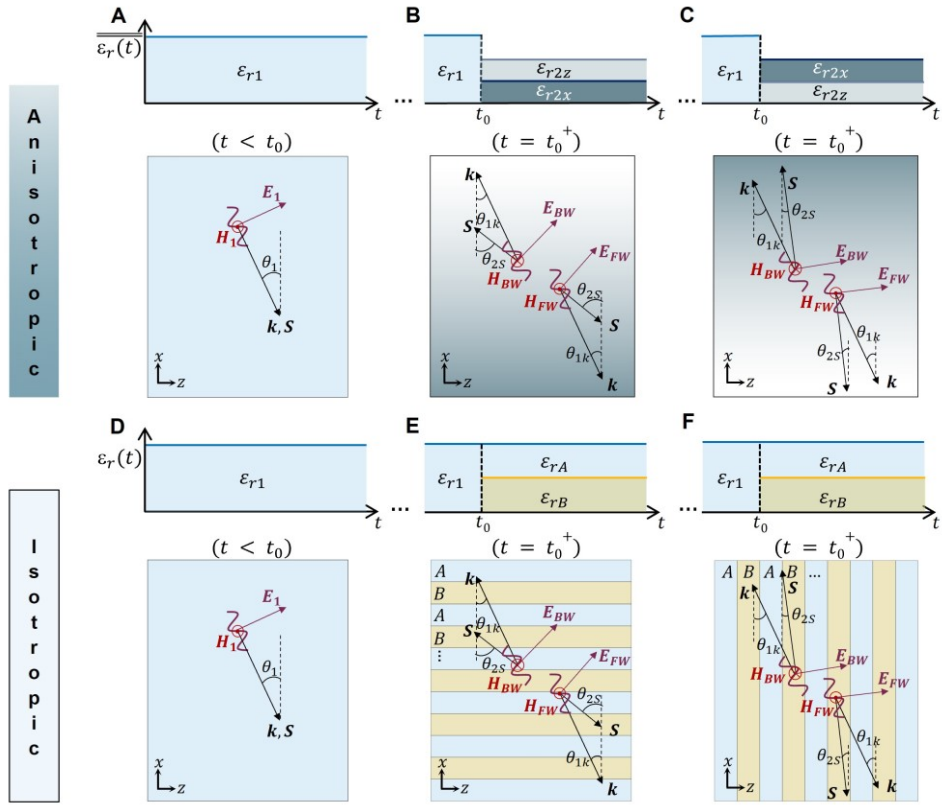

**Figure 1:** Schematic representations of an EM wave propagating in an unbounded, isotropic medium for $t < t_0$ (A). Then at $t = t_0$ an isotropic-to-anisotropic time interface is induced everywhere in space, resulting in a medium described by an anisotropic tensor, $\overline{\overline{\varepsilon_{r2}}}$, for $t > t_0$ with components (B) $\varepsilon_{r2x} < \varepsilon_{r2z}$ or (C) $\varepsilon_{r2x} > \varepsilon_{r2z}$. (D) a wave propagates in the same medium as (A). At $t = t_0$, isotropic-to-isotropic time interfaces are induced within discrete regions in space, resulting in a subwavelength periodic multilayer structure that has their layers (E) horizontally (parallel to $z$) or (F) vertically (parallel to $x$) aligned. The scenarios from (E,F) are the isotropic-to-isotropic spacetime equivalent of those from (B,C), respectively.

Once the effective performance of the scenarios from Fig. 1e,f is understood, we can study the effect of the resulting permittivity tensor $\overline{\overline{\varepsilon_{reff}}} = \{\varepsilon_{reff,x}, \varepsilon_{reff,z}\}$ when considering oblique incidence of the EM signal. As described above, and shown in Fig. 1a,d, the EM wave propagates within an isotropic, unbounded medium with an angle $\theta_1$ for $t < t_0$ (i.e., $\theta_1 = \theta_{2k} = \theta_{2S}$). Then, for $t > t_0$, as the multilayers are created simultaneously in time, the medium will now be described by an effective anisotropic tensor meaning that the direction of the Poynting vector will change (similar to the scenarios from Fig. 1b,c). To calculate $\theta_{2S}$, one can use the expression derived in[70] considering that the components of the relative permittivity tensor of the medium for $t > t_0$ are defined as in Eqs. 1,2 for the horizontal and vertical subwavelength periodic spatial multilayers respectively [i.e., the tensor is now defined in terms of $\varepsilon_{rA}, \varepsilon_{rB}$ and $\widetilde{\Delta x_A}$ (or $\widetilde{\Delta z_A}$)]. With this setup, the new direction of the energy propagation (Poynting vector) for $t > t_0$ considering the scenarios from Fig. 1e,f can be defined by the following expressions, respectively (please see full derivation in the Supplementary Materials document):

$$\theta_{2S} = \arctan\left\{tan(\theta_1)\left[\left(\frac{\varepsilon_{rA}}{\varepsilon_{rB}} + \frac{\varepsilon_{rB}}{\varepsilon_{rA}} - 2\right)\left(\widetilde{\Delta x_A} - \widetilde{\Delta x_A^2}\right) + 1\right]\right\} \quad (3a)$$

$$\theta_{2S} = \arctan\left[\frac{tan(\theta_1)}{\left(\frac{\varepsilon_{rA}}{\varepsilon_{rB}} + \frac{\varepsilon_{rB}}{\varepsilon_{rA}} - 2\right)\left(\widetilde{\Delta z_A} - \widetilde{\Delta z_A^2}\right) + 1}\right] \quad (3b)$$

As observed in the expressions from Eq. 3, the new directions of the Poynting vector, $\theta_{2S}$, have been rewritten in such a way that it only depends on the isotropic relative permittivity values of the temporally induced spatial layers ($\varepsilon_{rA,B}$) and the filling fraction (or geometrical dimensions of the spatial layers). i.e., Eq. 3 represents a fundamental result in our work as it demonstrates that isotropic-to-anisotropic time interfaces (Fig. 1a-c) can be emulated with isotropic-to-isotropic spacetime modulation instead (Fig. 1d-f). Finally, and for the sake of completeness, the amplitude of the induced FW and BW waves created by both scenarios from Fig. 1e,f can also be derived as the result of creating a spacetime effective medium. Following the same process as that of Eq. 3, the amplitude of the FW ($E_2^+/E_1$) and BW ($E_2^-/E_1$) waves can be defined as follow (see full derivation in the Supplementary Materials document):

$$\frac{E_2^\pm}{E_1} = \frac{1}{2}\left[\varepsilon_{r1}\sqrt{\left(\frac{C_H^2}{A^2} - \frac{1}{B_H^2}\right)sin^2(\theta_1) + \frac{1}{B_H^2}} \pm \sqrt{\varepsilon_{r1}}\sqrt{\frac{(B_H^2 C_H^2 - A^2)sin^2(\theta_1) + A^2}{(AB_H^2 C_H - A^2 B_H)sin^2(\theta_1) + A^2 B_H}}\right] \quad (4a)$$

$$\frac{E_2^\pm}{E_1} = \frac{1}{2}\left[\varepsilon_{r1}\sqrt{\left(\frac{1}{B_V^2} - \frac{C_V^2}{A^2}\right)sin^2(\theta_1) + \frac{C_V^2}{A^2}} \pm \sqrt{\varepsilon_{r1}}\sqrt{\frac{(A^2 - B_V^2 C_V^2)sin^2(\theta_1) + B_V^2 C_V^2}{(A^2 B_V - AB_V^2 C_V)sin^2(\theta_1) + AB_V^2 C_V}}\right] \quad (4b)$$

where $A = (\varepsilon_{rA}\varepsilon_{rB})$, $B_H = [\varepsilon_{rA}\widetilde{\Delta x_A} + \varepsilon_{rB}(1 - \widetilde{\Delta x_A})]$, $C_H = [\varepsilon_{rA}(1 - \widetilde{\Delta x_A}) + \varepsilon_{rB}\widetilde{\Delta x_A}]$, $B_V = [\varepsilon_{rA}\widetilde{\Delta z_A} + \varepsilon_{rB}(1 - \widetilde{\Delta z_A})]$ and $C_V = [\varepsilon_{rA}(1 - \widetilde{\Delta z_A}) + \varepsilon_{rB}\widetilde{\Delta z_A}]$, with subscripts $\{H,V\}$ presenting the temporally induced horizontal and vertical multilayers, respectively. Note that Eq. 4 considers the case when $\mu_{r1} = \mu_{r2y}$ but a general expression without this assumption is found in the Supplementary Materials document.

To study the implications of Eqs. 3,4, we provide specific examples in Fig. 2. Let us first discuss the scenario from Fig. 2a-d. Representing the case from Fig. 1d,e, it is considered that horizontal spatially subwavelength periodic multilayers are induced simultaneously in time via isotropic-to-isotropic time interfaces at $t = t_0$ (see the schematic representation of the time-dependent permittivity in Fig. 2a). With this configuration, we can start by defining a representative example for the scenario shown in Fig. 1a,b with $\varepsilon_{r1} = 10$ for $t < t_0$ and then induce in time an anisotropic permittivity tensor (via an isotropic-to-anisotropic time interface applied everywhere in space) with $\varepsilon_{r2x} = 8, \varepsilon_{r2z} = 12$ $(\mu_{r1} = \mu_{r2y})$ (it is important to note that while the values are large, these have been chosen as representative values taken from[70] to illustrate how the proposed spacetime configuration can mimic isotropic-to-anisotropic time interfaces). The idea here is to be able to emulate the same response but now with the configuration from Fig. 2a instead.

To do this, the first aspect one should enable is the emulation of the tensor $\overline{\overline{\varepsilon_{r2}}} = \{\varepsilon_{r2x} = 8, \varepsilon_{r2z} = 12\}$. For this, we consider the following parameters: a filling fraction of $\widetilde{\Delta x_A} = 0.5$, layers A with $\varepsilon_{rA} = \varepsilon_{r1}$ (i.e., time-independent) whereas for layers B, $\varepsilon_{rB} \neq \varepsilon_{r1}$. Using Eq. 1 with $\varepsilon_{reff,x} = 8$ and $\varepsilon_{reff,z} = 12$ to mimic the isotropic-to-anisotropic time interface, one can calculate the relative isotropic permittivity values for layers A and B. However, due to the geometry of the induced spatial multilayers in time (horizontal in this case) and because the permittivity of one of the layers is time-invariant (layer A with $\varepsilon_{rA} = \varepsilon_{r1}$), both components of the effective tensor $\overline{\overline{\varepsilon_{reff}}}$ will always be larger or smaller than $\varepsilon_{r1}$. This means that the condition $\varepsilon_{reff,x} < \varepsilon_{r1} < \varepsilon_{reff,z}$ required to mimic the target homogeneous case described above $[(\varepsilon_{r2x} = 8) < (\varepsilon_{r1} = 10) < (\varepsilon_{r2z} = 12)]$ cannot be fulfilled. However, one can recalculate $\varepsilon_{r1}$ for the induced multilayer structure in time using Eq. 1 with $\varepsilon_{rA} = \varepsilon_{r1}$ and $\varepsilon_{rB} \neq \varepsilon_{r1}$. In doing so, $\overline{\overline{\varepsilon_{reff}}} = \overline{\overline{\varepsilon_{r2}}}$ can still be satisfied even when $\varepsilon_{reff,x} < \varepsilon_{r1} < \varepsilon_{reff,z}$ is not fulfilled. In this case, using Eq. 1, the resulting value is $\varepsilon_{r1} = 18.93$ for the structure from Fig. 2a. To illustrate these results, the $x$ and $z$ components of $\overline{\overline{\varepsilon_{reff}}}$ for the structure from Fig. 2a as a function of the ratio between the isotropic relative permittivity of layers B and A are shown in Fig. 2b using $\varepsilon_{r1} = \varepsilon_{rA} = 18.93$. As observed, when $\varepsilon_{rB}/\varepsilon_{rA} = 1$, $\varepsilon_{reff,x} = \varepsilon_{reff,z} = \varepsilon_{r1} = 18.93$, as expected given that no spatial multilayers are induced in time. For values of $\varepsilon_{rB} < \varepsilon_A$ or $\varepsilon_{rB} > \varepsilon_A$, the components of the tensor are both smaller or larger than $\varepsilon_{r1} = \varepsilon_{rA}$, respectively, which corroborates the need of redefining $\varepsilon_{r1}$, as discussed above. In this figure, the black dashed line represents the case where $\varepsilon_{reff,x} = 8, \varepsilon_{reff,z} = 12$, which is achieved when $\varepsilon_{rB} = 5.07$. This means that, based on the effective permittivity tensor alone, it is possible to mimic the homogeneous tensor without inducing isotropic-to-anisotropic time interfaces.

To fully demonstrate that the structure from Fig. 2a can emulate the homogeneous scenario from Fig. 1a,b, the theoretical results of the direction of the Poynting vector ($\theta_{2S}$), for times $t > t_0$, as a function of the incident angle ($\theta_1$) using $\varepsilon_{rA} = \varepsilon_{r1} = 18.93$, $\varepsilon_{rB} = 5.07$ and $\widetilde{\Delta x_A} = 0.5$ with Eq. 3a are shown in Fig. 2c as symbols together with the theoretical results for the homogeneous case[70], with $\varepsilon_{r2x} = 8, \varepsilon_{r2z} = 12$, as solid lines. The agreement between the results is clear, demonstrating the potential of the proposed isotropic-to-isotropic spacetime structure to emulate an isotropic-to-anisotropic time interface. However, it is important to note that even when the angle of the energy propagation ($\theta_{2S}$) is the same for both the multilayer and the homogeneous cases, the amplitude of the FW and BW waves may differ as shown in Fig. 2d. In these results, the multilayer scenario (symbols) perfectly matches that of the homogeneous case (solid lines) when $\varepsilon_{r1} = 18.93$, but differ for the homogeneous case from[70] where $\varepsilon_{r1} = 10$. This result is as expected and occurs because of the need of modifying $\varepsilon_{r1}$ for times $t < t_0$ when one of the layers is time-invariant (as explained above). To overcome this, one can simply allow for both layers A and B to change in time by inducing two different isotropic-to-isotropic time interfaces simultaneously. This latter scenario is schematically shown in Fig. 2e. To analyse this case, at $t = t_0$ isotropic-to-isotropic time interfaces are induced in discrete spatial regions to create spatial layers (filling fraction of $\widetilde{\Delta x_A} = 0.5$) with $(\varepsilon_{rA}, \varepsilon_{rB}) \neq \varepsilon_{r1} = 10$. i.e., we can force $\varepsilon_{r1} = 10$ as in[70].

As an example, we can try to mimic an isotropic-to-anisotropic time interface with parameters $\varepsilon_{r2x} = 12$, $\varepsilon_{r2z} = 20$ using the spacetime structure from Fig. 2e. Using Eq. 1, the theoretical results of the resulting components of the permittivity tensor for the multilayer case are shown in Fig. 2f as a function of the isotropic permittivity values for layers A and B. From these results, one can map the required values for $\varepsilon_{rA,B}$ to obtain the effective values of the tensor required (i.e. $\varepsilon_{r2x} = 12$, $\varepsilon_{r2z} = 20$), resulting in $\varepsilon_{rA} = 32.65$ and $\varepsilon_{rB} = 7.35$. With these parameters, the calculated Poynting vector (Eq. 3a) and the amplitudes of the FW and BW waves (Eq. 4a) as a function of $\theta_1$ are shown as symbols in Fig. 2g,h. For completeness, the theoretical calculations of the same parameters for the homogeneous isotropic-to-anisotropic time interface are also shown in the same panels as solid lines, demonstrating an agreement between them.

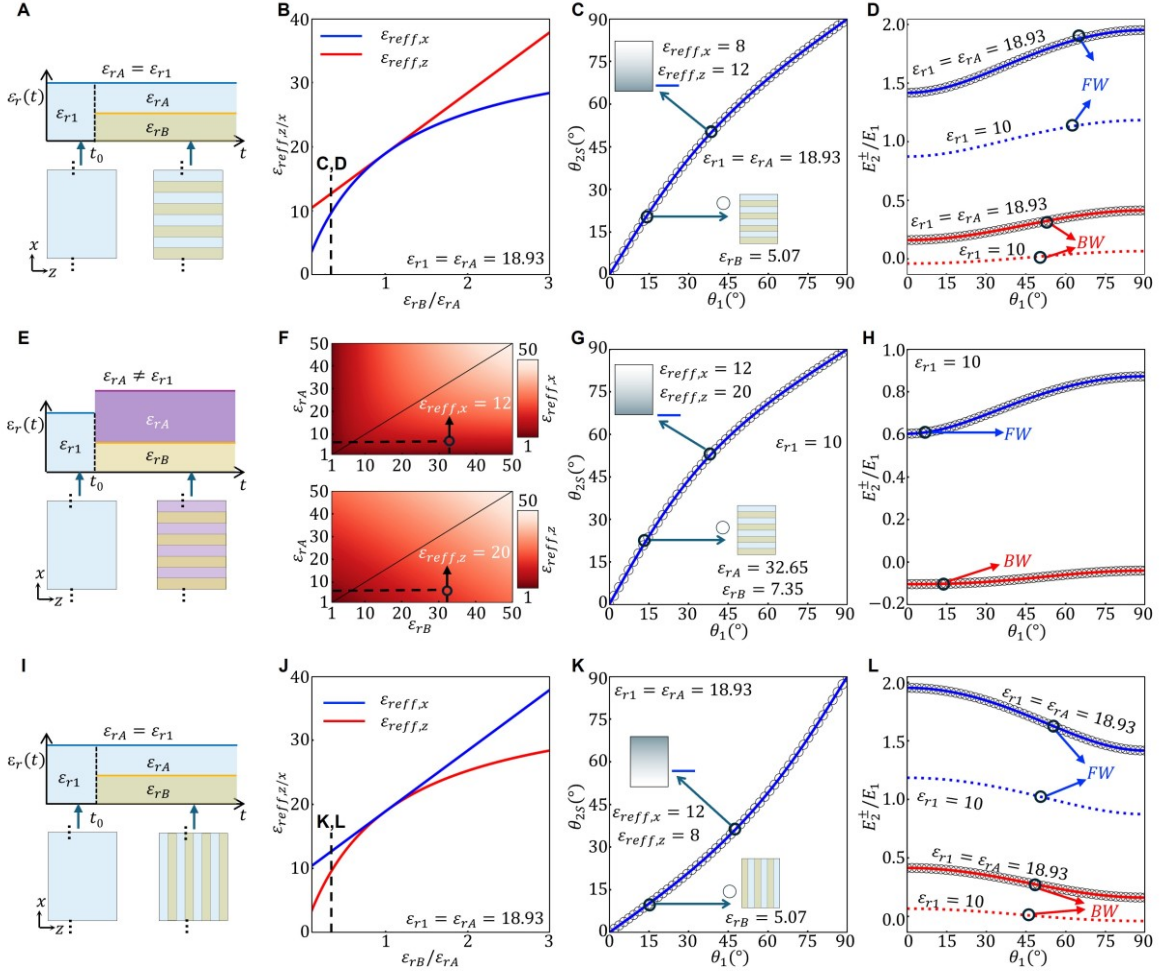

**Figure 2:** (A) Schematic of the time-dependent $\varepsilon_r$ for the scenario from Fig. 1d,e where $\varepsilon_{r1} = \varepsilon_{rA}$. (B) Effective values of the permittivity tensor $\varepsilon_{reff,x}$ (blue) and $\varepsilon_{reff,z}$ (red) as a function of the ratio between the isotropic permittivities of the induced spatial layers ($\varepsilon_B/\varepsilon_A$) for $\varepsilon_{r1} = \varepsilon_{rA} = 18.93$ with a filling fraction of 0.5. (C,D) Direction of Poynting vector and amplitude of the FW and BW waves, respectively, as a function of the incident angle ($\theta_1$) for the scenario from (A) (black circles) along with the results for the homogeneous isotropic-to-anisotropic time interface equivalent (blue line). The dotted lines in (D) correspond homogeneous isotropic-to-anisotropic time interface equivalent when $\varepsilon_{r1} = 10$ while the solid lines assume $\varepsilon_{r1} = 18.93$. (E, G, H), similar to (A,C,D) but for the case when both induced periodic subwavelength multilayers are time-dependent; i.e., $(\varepsilon_{rA}, \varepsilon_{rB}) \neq \varepsilon_{r1}$. (F) Components of the effective permittivity tensor for the scenario from (E) as a function of the isotropic relative permittivity of layers A and B. (I-L) Same configuration as in (A-D) but for the case where vertical multilayers are induced in time at $t = t_0$.

Finally, and to fully study the scenarios proposed in Fig. 1d-f, we present the results when vertical subwavelength periodic multilayers are induced in time via simultaneous time interfaces created at $t = t_0$ in Fig. 2i-l. As observed, this configuration is complementary to that discussed in Fig. 2a-d, with the components of the effective permittivity tensor exchanged (as expected due to the form of Eq. 2a,b). The results discussed from Fig. 2 show that the isotropic permittivity values that make up the spacetime structures can be used to engineer wave behaviour in a similar way as isotropic-to-anisotropic time interfaces. To support the results discussed in this section, numerical simulations are presented below to further corroborate the potential of isotropic-to-isotropic spacetime interfaces to mimic isotropic-to-anisotropic time interfaces.

## 2.2 Numerical results isotropic-to-isotropic spacetime

To begin, let us evaluate the scenarios from Fig. 1b,e where the permittivity of the medium is changed from isotropic to anisotropic in time (Fig. 1b, homogeneous scenario) or when a spatially subwavelength multilayer structure is induced in time via multiple simultaneous time interfaces (Fig. 1e). As explained above, the effective medium from Fig. 1e will create an effective permittivity tensor satisfying $\varepsilon_{reff,x} < \varepsilon_{reff,z}$. With this in mind, the numerical results of the out-of-plane magnetic field distribution $H_z$ of an EM wave propagating with an incident angle of $\theta_1 = 25°$ in an isotropic, homogeneous medium for a time $t < t_0$ is shown in Fig. 3a considering $\varepsilon_{r1} = 10$ (the relative permeability is time-independent with $\mu_{r1} = \mu_{r2y} = 1$). For this time, and for all $t < t_0$, the wavenumber and the direction of energy propagation are aligned, as expected. Then, at $t = t_0 = 12T_1$ [with $T_1 = 0.67$ ns as the period of the incident monochromatic signal for times $t < t_0$ ($f_1 = 1.5$ GHz as an example)] simultaneous isotropic-to-isotropic time interfaces are created within different regions in space such that a subwavelength periodic multilayer structure is created with $\varepsilon_{rA} = \varepsilon_{r1} = 10$ and $\varepsilon_{rB} = 1$ (see schematic representations on the top row of Fig. 3). As shown in Fig. 3b,c (top), the spatial period of the multilayer structure is $0.2\lambda$ along the $x$ axis (with $\lambda$ as the wavelength of the incident wave) with a layer thickness of $0.1\lambda$ (i.e., a filling fraction of $\widetilde{\Delta x_A} = \widetilde{\Delta x_B} = 0.5$). With this configuration, the numerical results of the $H_y$ for times $t > t_0$ are shown in Fig. 3b,c (specifically, at $t = t_0^+ = 15T_1$ and $t > t_0^+ = 16.5T_1$, respectively). As observed, the FW and BW waves are created, with their direction no longer being aligned to $\theta_1$ (dashed line). These results demonstrate that, indeed, it is possible to mimic isotropic-to-anisotropic time interfaces using isotropic-to-isotropic spacetime modulations. Quantitatively, using Eq. 1 and Eq. 3a, the effective permittivity tensor and the direction of the Poynting vector for $t > t_0$ is $\varepsilon_{reff,x} = 1.82 < \varepsilon_{reff,z} = 5.5$ and $\theta_{2S} = 54.67°$, respectively. To guide the eye, the arrows representing the FW and BW waves from Fig. 3b,c have been aligned to this angle $\theta_{2S}$, showing how the simulation results are in agreement with the theoretical predictions.

To further analyse the results for the isotropic-to-isotropic spacetime scenario, the numerical results of the out-of-plane magnetic field distribution for the homogeneous configuration calculated at the same times as those from Fig. 3a-c are presented in Fig 3d-f, respectively. Here, the same EM wave as in Fig. 3a propagates within the unbounded medium for times $t < t_0$. Then, at $t = t_0$ an isotropic-to-anisotropic time interface is induced everywhere in space resulting in a homogeneous anisotropic medium described by a relative permittivity tensor, $\overline{\overline{\varepsilon_{r2}}} = \{\varepsilon_{r2x} = 1.82, \varepsilon_{r2z} = 5.5\}$. From these results, the agreement with the multilayer structure induced in time is clear (see Fig. 3b,c and Fig. 3e,f). To better compare these results, the theoretical calculations (using[70]) of the out-of-plane magnetic field distribution before and after inducing the time interface for the homogeneous medium are shown as insets in Fig. 3d-f. Here, the theoretical results of the Poynting vector are also shown (arrows within the insets). By inspecting these results, both the Poynting vector and the amplitudes of the FW and BW waves are in agreement with the numerical results with $\theta_{2S} = 54.67°$. It is interesting to note how for the proposed horizontal subwavelength periodic multilayer structure induced in time (Fig. 3a-c) the direction of the Poynting vector is always $\theta_{2S} > \theta_1$. This is theoretically shown in Fig. 3h, where the analytical values of $\theta_{2S}$ as a function of $\varepsilon_{rB}/\varepsilon_{rA}$ are presented for different incident angles (namely $\theta_1 = 25°, \theta_1 = 45°$ and $\theta_1 = 65°$) (the vertical grey line in this panel represents the case when $\varepsilon_{rB}/\varepsilon_{rA} = 1$ so $\theta_1 = \theta_{2S}$). This response, however, will change when inducing vertical subwavelength spatial multilayers, as it will be shown later.

Finally, and to fully analyse the potential of the configuration from Fig. 1d,e to emulate an isotropic-to-anisotropic time interface, we recorded the $H_y$ at a fixed position in space as a function of time and the results are shown in Fig. 3g for both cases from Fig. 3a-c and Fig. 3d-f (red and black lines, respectively) (the exact position in space is represented as a circle in Fig. 3a,d). As observed, a good agreement is obtained between the results, demonstrating how the multilayer configuration has some "ripples" due to the induced spatial multilayer in time. These ripples will transform into higher-order harmonics which can be removed using filters (as suggested in[58,59], outside of the scope of this manuscript). As expected, increasing the number of layers per wavelength would decrease these ripples (as it will be shown for the vertical multilayer configuration below). However, it can be clearly seen in Fig. 3g that $H_y$ for both the isotropic-to-isotropic spacetime and the isotropic-to-anisotropic time interface has the same period for times $t > t_0$ (i.e., the same frequency). As it is known, at a time interface the frequency is modified but the wavenumber (and wavelength) is preserved. From[57,70], for an isotropic-to-anisotropic time interface, the new frequency is $f_2 = \left(\frac{1}{2\pi}\right) c \sqrt{\left[\frac{k_x^2}{\varepsilon_{r2z}\mu_{r2y}}\right] + \left[\frac{k_z^2}{\varepsilon_{r2x}\mu_{r2y}}\right]}$, where $c$ is the velocity of light in a vacuum, $k_x = -k\cos(\theta_1)$, $k_z = k\sin(\theta_1)$ and $k = \frac{\omega_1}{v_1}$, $v_1 = \frac{c}{\sqrt{\mu_{r1}\varepsilon_{r1}}}$. From this expression, the predicted frequency for the isotropic-to-anisotropic case from Fig. 3d-f is $f_2 = 1.573f_1$. For the scenario from Fig. 3a-c, $f_2 \approx 1.573f_1$ (calculated by measuring the period of the red curve from Fig. 3g), demonstrating an excellent agreement with the theoretical calculations of the homogeneous scenario (see Supplementary video 1 for an animation of the results shown in Fig. 3a-f).

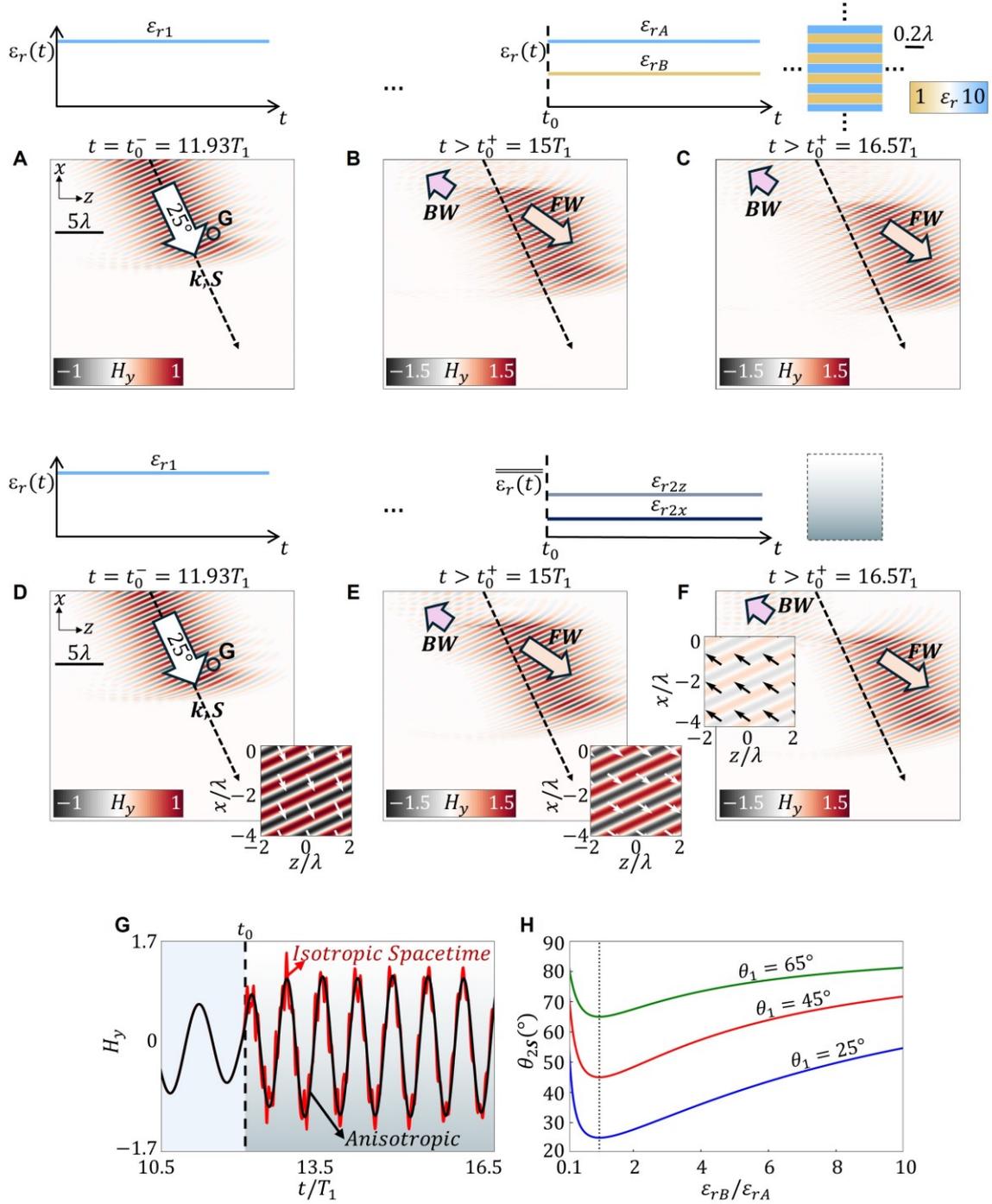

**Figure 3:** Numerical simulation results showing the out-of-plane magnetic field, $H_y$, for an oblique ($\theta_1 = 25°$) incident EM wave $f_1 = 1.5$ GHz propagating in (A,D) an initial isotropic, unbounded, medium with $\varepsilon_{r1} = 10$. Then at $t = t_0 = 12T_1$ (B-C) isotropic-to-isotropic spacetime interfaces are induced simultaneously to create a horizontal subwavelength periodic multilayer structure with $\varepsilon_{rA} = \varepsilon_{r1} = 10$ and $\varepsilon_{rB} = 1$ (see top insets) or (E,F) an isotropic-to-anisotropic time interface is induced everywhere in space with $\varepsilon_{r2} = [\varepsilon_{r2x} = 1.82, \varepsilon_{r2z} = 5.5]$. The insets in (D,E,F) represent the theoretical calculations of $H_y$ showing the Poynting vector as arrows. (G) Out-of-plane $H_y$ for the scenarios shown in (A-C) and (D-F) (red and black lines, respectively) as a function of time recorded at a single point in space (marked as circles in A,D). The time interface at $t = t_0$, is noted by the black dashed line. (H) Theoretical $\theta_{2S}$ as a function of $\varepsilon_{rB}/\varepsilon_{rA}$ for different values of $\theta_1$ using $\varepsilon_{rA} = 10$ and $\widetilde{\Delta x_{A,B}} = 0.5$ respectively. The case where $\varepsilon_{rA} = \varepsilon_{rB}$ is indicated by the grey dotted line.

The agreement between both isotropic-to-isotropic spacetime modulations and isotropic-to-anisotropic time interfaces for the homogeneous case represents a key result in this work, as one can verify that it is possible to emulate temporal anisotropy using isotropic modulations in spacetime. To fully explore the scenarios from Fig. 1, let us now briefly study the case shown in Fig. 1d,f where a vertical subwavelength multilayer structure is induced in time via simultaneous isotropic-to-isotropic time interfaces. Here, we start with the same isotropic, homogeneous medium from Fig. 3a,d for times $t < t_0$ considering that $\varepsilon_{r1} = 10$, $\mu_{r1} = \mu_{r2y} = 1$ and $\theta_1 = 25°$ for the incident EM wave. Then, at $t = t_0$, multiple isotropic-to-isotropic time interfaces are induced simultaneously creating the multilayer structure from Fig. 4a,b (top panels) with $\varepsilon_{rA} = \varepsilon_{r1} = 10$, $\varepsilon_{rB} = 1$ and $\widetilde{\Delta x_{A,B}} = 0.5$ and a spatial periodicity of $0.2\lambda$ along the $z$ axis. With these parameters, and using Eq. 2, the permittivity tensor emulated by this structure is $\overline{\overline{\varepsilon_{reff}}} = \{\varepsilon_{reff,x} = 5.5, \varepsilon_{reff,z} = 1.82\}$ (i.e., $\varepsilon_{reff,x} > \varepsilon_{reff,z}$, as represented in Fig. 1c,f). Interestingly, using this configuration along with Eq. 3b, the theoretical direction of the Poynting vector for $t > t_0$ is $\theta_{2S} = 8.76°$, meaning that $\theta_{2S} < \theta_1$ (which is the complementary performance to the horizontal multilayer configuration described in Fig. 3, as explained above) (see also the Supplementary materials for further examples). With this configuration, the numerical results of the out-of-plane $H_y$ distribution for times $t > t_0$ using the vertical multilayer configuration induced in time are shown in Fig. 4a,b (the results for the homogeneous configuration using an isotropic-to-anisotropic time interface are shown in the Supplementary materials document). As observed, the direction of the energy propagation is now $\theta_{2S} < \theta_1$ in agreement with the theoretical predictions. To further analyse these results, the recorded $H_y$ at a single position in space as a function of time for both the vertical multilayer configuration and the homogeneous case are shown as blue and black lines on the top panel of Fig. 4c. From these results, it is clear that the frequency is modified when inducing the time interfaces with the theoretically predicted frequency as $f_2 = 2.2f_1$, in agreement with the one calculated from the numerical results using the homogeneous configuration (black line) $f_2 \approx 2.2f_1$. However, the frequency for the vertical multilayer scenario is different. This can be further evidenced by looking at the spectrum shown on the bottom panel from Fig. 4c (calculated from the temporal signals). This is because the effective medium theory in space is not fulfilled for a period of $0.2\lambda$ along the $z$ axis using a vertical multilayer structure (note that even when $\theta_1$ is the same as the one used for the configuration from Fig. 3, for the vertical multilayer the *effective* thickness along the direction $\theta_1$ is larger).

To verify this, the numerical results of $H_y$ (considering that a vertical subwavelength multilayer structure is created in time with a spatial period of $0.1\lambda$ along the $z$ axis) are shown in Fig. 4d-f. From these results, it is now clear how reducing the spatial period improves the agreement with the theoretical predictions of an isotropic-to-anisotropic time interface in a homogeneous medium. Moreover, note that the "ripples" observed in Fig. 3 are substantially reduced in the case from Fig. 4d-f given that the multilayer structure closely resembles the homogeneous scenario, as discussed before. For completeness, an animation showing the results discussed in Fig. 4d-f is shown in Supplementary video 2.

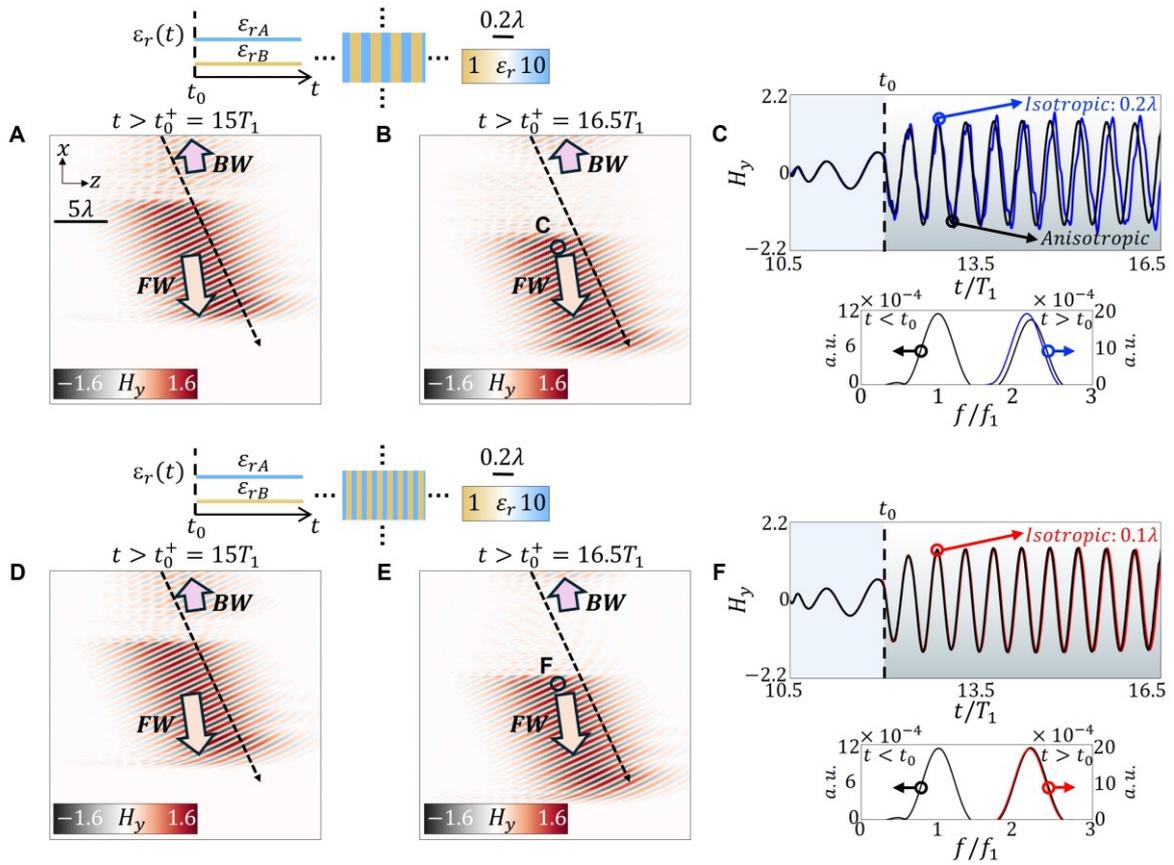

**Figure 4:** Numerical results of the $H_y$ distribution at (A,D) $15T_1$ and (B,E) $16.5T_1$ for the case when a vertical subwavelength periodic multilayer is induced in time (with $\varepsilon_{rA} = \varepsilon_{r1} = 10, \varepsilon_{rB} = 1$ and $\widetilde{\Delta z_{A,B}} = 0.5$) using a spatial period along the $z$ axis of (A,B) $0.2\lambda$ and (D,E) $0.1\lambda$. The black dashed arrow represents $\theta_1 = 25°$. (C,F) Recorded $H_y$ as a function time (top), extracted at a single spatial position (marked as a circle in B,E) with the blue/red lines representing the spacetime scenarios (isotropic-to-isotropic spacetime) and the black lines as the results for the homogeneous isotropic-to-anisotropic time interface equivalent. The bottom panels from (C,F) show the spectral content of the temporal signals from the top panels.

## 2.3 The effect of the spatial filling fraction.

The results discussed in the previous sections considered the case where the layers A and B of the induced subwavelength periodic multilayer structures had equal thickness, i.e., $\widetilde{\Delta x_{A,B}}$ (or $\widetilde{\Delta z_{A,B}}$) = 0.5. Here, we discuss the effect when the filling fraction is changed. First, let us consider the same isotropic medium previously discussed in Fig. 3a,d (for times $t < t_0$ $\varepsilon_{r1} = 10$, $\mu_{r1} = \mu_{r2y} = 1$ and $\theta_1 = 25°$ for the incident EM signal). Then at $t = t_0$, multiple isotropic-to-isotropic time interfaces are induced to form a horizontal (vertical) subwavelength multilayer structure as the one shown in Fig. 1b,e with $\varepsilon_{rA} = \varepsilon_{r1} = 10$, $\varepsilon_{rB} = 2$, and $\widetilde{\Delta x_A} \neq 0.5$ ($\widetilde{\Delta z_A} \neq 0.5$). With this configuration and using Eqs. 1-2 and Eq. 3, the theoretical results of the two components of the effective permittivity tensor $\varepsilon_{reff,x,z}$ and the direction of the Poynting vector ($\theta_{2S}$), respectively, for times $t > t_0$ as a function of the filling fraction of layers A are shown in Fig. 5a (with the top and bottom panels showing the values for the horizontal and vertical multilayer structures, respectively). These results further corroborate the conditions mentioned in the previous section where the resulting direction of the Poynting vector fulfils $\theta_{2S} > \theta_1$ and $\theta_{2S} < \theta_1$ for the horizontal and vertical multilayer configuration, respectively. Moreover, note that $\theta_{2S}$ can change depending on the filling fraction, an interesting property to produce different responses using the same materials but with different thicknesses. As observed, when the filling fractions are 0 or 1 $\theta_{2S} = \theta_1$, meaning that the isotropic-to-isotropic time interface is applied to the entire region in space (i.e., a homogeneous scenario). Interestingly, note how $|\theta_{2S} - \theta_1|$ is maximized when $\widetilde{\Delta x_A}, \widetilde{\Delta z_A} = 0.5$. At this filling fraction, the difference between the two effective components $\varepsilon_{reff,x}, \varepsilon_{reff,z}$ is also maximized.

To numerically evaluate the results from Fig. 5a, we provide in Fig. 5b,c the numerical results of the out-of-plane magnetic field for times $t > t_0$ considering the case when a horizontal multilayer structure is created in time with $\widetilde{\Delta x_A} = 0.25$ (see the schematic of the structure on the top panel). Using Eq. 1 and Eq. 3a, the components of the permittivity tensor are $\varepsilon_{reff,x} = 2.5, \varepsilon_{reff,z} = 4$ and $\theta_{2S} \approx 36.73°$, respectively. These results are in agreement with the homogeneous isotropic-to-anisotropic equivalent (see Fig. 5e,f). To further analyse these results, the recorded $H_y$ as a function of time along with the spectral response for both scenarios from Fig. 5b,c and Fig. 5e,f are shown on the top and bottom panels from Fig. 5d, respectively, showing a good agreement with the new frequency for times $t > t_0$ as $f_2 = 1.664 f_1$. To finalize, as discussed in the introduction, the scientific community is working towards exploring different experimental approaches to control waves both in space and time. Examples include transmission lines and thin materials such as ITO as a platform for time-varying applications[93,95,98,99], among others. We envision that the temporally induced multilayer structures proposed in our work will leverage these and other efforts from the scientific community as a mechanism for its experimental demonstration.

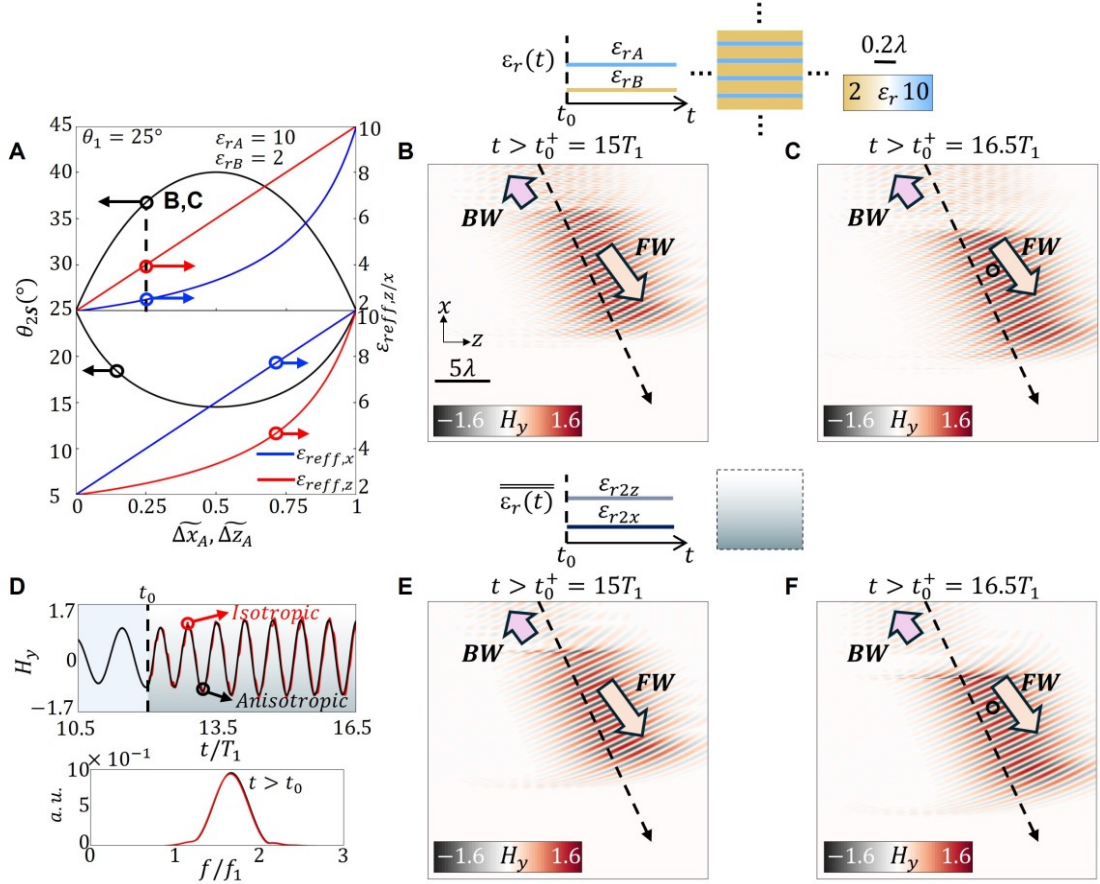

**Figure 5:** (A) Analytical results for $\theta_{2S}$ (black line) and $\varepsilon_{reffx,z}$ (blue and red lines, respectively) as a function of $\widetilde{\Delta x_A}$ (or $\widetilde{\Delta z_A}$), for the horizontal (vertical) multilayer configurations, respectively (top and bottom panels). Numerical results of the out-of-plane magnetic field, $H_y$, distribution at (B,E) $t = 15T_1$ and (C,F) $t = 16.5T_1$ for the case where a horizontal multilayer is induced in time with a filling fraction $\widetilde{\Delta x_A} = 0.25$ and $\varepsilon_{rA} = 10, \varepsilon_{rB} = 2$ (B,C) and where the equivalent isotropic-to-anisotropic time interface is induced in time everywhere in space (E,F). (D) $H_y$ recorded at a single spatial point (top panel), marked by a black circle in panels (C,F) with red and black lines representing the results for the case using the induced spacetime multilayer structure via isotropic-to-isotropic time interfaces and when considering an isotropic-to-anisotropic time interface in an homogenous, unbounded medium, respectively. The bottom panel from (D) represents the spectral content.

## 3. Conclusions

In this work, the possibility of emulating photonic isotropic-to-anisotropic time interfaces has been proposed and demonstrated using isotropic-to-isotropic spacetime modulations. To do this, it was considered that the permittivity of an unbounded, isotropic, homogeneous medium was changed in time by inducing isotropic-to-isotropic time interfaces in discrete regions in space to create subwavelength spatially periodic multilayers. In so doing, the system was modelled as an effective permittivity tensor after the time interfaces were introduced. A full theoretical analysis has been presented with all the results being supported by numerical simulations, showing an agreement between them. Our results may open new direction for implementation of isotropic-to-anisotropic time interfaces using isotropic-to-isotropic spacetime modulations inspired by spacetime effective medium approaches.

## 4. Methods

The numerical results presented in this work were carried out using the time domain solver of the commercial software COMSOL Multiphysics®. For the isotropic-to-anisotropic time interfaces where the time modulated permittivity is applied everywhere in space, the same configuration as in[70] was used. The simulations of an induced horizontal (vertical) subwavelength periodic multilayer structure shown in Fig. 2-5 were carried out considering a monochromatize EM wave being excited by a scattering boundary condition placed on the top boundary of an $25\lambda \times 35\lambda$ simulation box (along the $x$ and $z$ axis, respectively). The monochromatic beam was calculated as in[86]. For the other three boundaries, scattering boundary condition were also used but with the "no incident field" option to avoid undesirable reflections. The incident signal was switched off at $t = t_0$ once time interfaces were applied. The subwavelength spatial multilayers were implemented using an analytical function considering a built-in step function with a transition time as in our previous works[86] and two continuous derivatives. A user-controlled mapped mesh was implemented to accurately simulate the subwavelength multilayers with a minimum/maximum mesh size of $7 \times 10^{-4}\lambda/0.05\lambda$ for the multilayers with a period of $0.2\lambda$ and $7 \times 10^{-4}\lambda/0.025\lambda$ for the simulations with a period of $0.1\lambda$. For the results from Fig. 5, where the filling fraction was modified, the same parameters as with a period of $0.1\lambda$ were used.

## Acknowledgements

This work was supported by the Leverhulme Trust under the Leverhulme Trust Research Project Grant scheme (RPG-2023-024). For the purpose of Open Access, the authors have applied a CC BY public copyright license to any Author Accepted Manuscript (AAM) version arising from this submission.

## Conflicts of interest

The authors declare no conflicts of interest.

## Data availability

The datasets generated and analysed during the current study are available from the corresponding author.